\documentclass[twocolumn,showpacs,nofootinbib]{revtex4-1}
\usepackage{epsfig,amssymb,amsmath,latexsym}
\usepackage{pifont}
\usepackage{hyperref}

\usepackage{color}
\definecolor{nicered}{rgb}{0.7,0.1,0.1}
\definecolor{nicegreen}{rgb}{0.1,0.5,0.1}

\newcommand{\be}{\begin{equation}}
\newcommand{\ee}{\end{equation}}
\newcommand{\bea}{\begin{eqnarray}}
\newcommand{\eea}{\end{eqnarray}}

\newcommand{\no}{\noindent}
\newcommand{\nb}{\nonumber}

\newcommand{\de}{\partial}

\renewcommand\l{\lambda}

\renewcommand\a{\alpha}

\newcommand\m{\mu}
\newcommand\n{\nu}

\newcommand\V{{\ensuremath{\cal V}}}
\newcommand\tr{\text{Tr}}

\renewcommand\l{\ensuremath{\lambda}}

\newcommand\ba{\begin{array}}
\newcommand\ea{\end{array}}

\newcommand\SEC[1]{\medskip\noindent{\sl\bfseries #1}}

\begin{document}

\title{New Branches of Massive Gravity}

\author{D. Comelli$^a$, M. Crisostomi$^b$, K. Koyama$^b$, L. Pilo$^{c,d}$ and G. Tasinato$^{e}$}
\affiliation{
   $^a$INFN, Sezione di Ferrara, I-35131 Ferrara, Italy\\
   $^b$Institute of Cosmology and Gravitation,~University of Portsmouth, Portsmouth, PO1 3FX, UK\\
   $^c$Dipartimento di Fisica, Universit\`a di L'Aquila, I-67010 L'Aquila, Italy\\
   $^d$INFN, Laboratori Nazionali del Gran Sasso, I-67010 Assergi, Italy\\
   $^e$Department of Physics, Swansea University, Swansea, SA2 8PP, UK
   }
\date{\small \today}

\begin{abstract}
\no The basic building block for Lorentz invariant and ghost free
massive gravity is the square root of the combination $g^{-1}\eta\,$, where $g^{-1}$ is the inverse of the physical metric and $\eta$ is a reference metric. Since the square root of a matrix is not uniquely defined, it is possible to have physically inequivalent 
potentials  corresponding  to  different branches.
We show that around Minkowski background the only perturbatively well defined branch is the potential proposed by de Rham, Gabadadze and Tolley.
On the other hand, if  Lorentz symmetry is broken spontaneously,
other potentials exist with a standard perturbative expansion.
We show this explicitly building new Lorentz invariant, ghost-free massive gravity potentials for theories that in the background preserve rotational invariance, but break Lorentz boosts.
\end{abstract}


\maketitle

\SEC{Introduction.} \;\;
Much  progress has been recently made in understanding massive
gravity.  Attempts to give a mass to the graviton
date back to the work of Fierz and Pauli (FP) in 1939 \cite{Fierz:1939ix}. They considered a mass term for a spin two field which is uniquely determined
requiring the absence of ghost degrees of 
freedom (dof). The mass term breaks the diffeomorphism invariance of General Relativity (GR), leading to a graviton with five degrees of freedom instead of the two of GR. In 1972, Boulware and Deser (BD) found that the  
ghostly sixth mode, removed by the FP tuning, reappears at non-linear level or around non-trivial backgrounds \cite{BD}. It is now well known that it is
possible to avoid the presence of the BD ghost by choosing a suitable
potential~\cite{Gabadadze:2011}. Such a ghost free theory can be
constructed by adding to the Einstein-Hilbert action a
potential $V$ function of the symmetric polynomials of the
square root of the matrix $X^{\mu}_{\,\,\nu} \equiv g^{\mu\a} f_{\a\nu}$, where $f_{\mu \nu}$ is a fixed reference 
metric~\cite{Hassan:2011vm}. In the following we will consider the
case  $f_{\mu \nu}= \eta_{\mu \nu}$, i.e. the reference metric
coincides with the Minkowski metric. This theory, dubbed de
Rham-Gabadadze-Tolley (dRGT) massive gravity, is Lorentz invariant (LI) in the unitary gauge\footnote{Invariance under coordinate transformation can be restored by introducing a
suitable set of St\"uckelberg fields ~\cite{HGS}.}.
More in general, the
whole class of rotationally invariant massive gravity theories with five degrees of freedom can be constructed~\cite{uscan,gencons} by using the canonical analysis, which can be extended to any non-derivative modified gravity theory~\cite{Comelli:2014xga}.

In this work we present alternative consistent branches for LI massive gravity. We show how their perturbative expansion and construction is related to the spontaneous
breakdown of the Lorentz symmetry. We also briefly comment on some of their phenomenological consequences.

\SEC{Ghost free potentials.} \;\;
From the canonical analysis of a generic, non-linear massive deformation of GR in four dimensions~\cite{uscan} 
\be
S =  \int d^4x \sqrt{g} M_{pl}^2 \left( R 
  - m^2 \, V \right) \equiv S_{EH} + S_V \,,
\label{act}
\ee
it follows that, in the unitary gauge, in order to have 5 propagating dof, the potential $V$ has to satisfy two conditions. In terms of the ADM variable $(N, N^i, \gamma_{ij})$ they read
\be 
{\rm Rank} |\V_{AB}|=3\,,
\qquad {\chi^0}^2 \,  \, {\cal V}_i  + 2  \, \chi^A  \chi^j
\,\frac{\de {\cal V}_A}{\de \gamma^{ij}} \,=0  \, ;  
\label{dini}
\ee
where we  define $\V\equiv  N \, V$.  $\V_A$ is the derivative
of $\V$ with respect to $N^A=N_A=(N, \,N^i)$ whose components are the
lapse and shifts;  $\chi^A$ is the eigenvector associated to the null eigenvalue of $\V_{AB}$. 
Requiring  that $V$ has a residual Lorentz invariance  besides rotational invariance forces the potential to be a
function of the eigenvalues~$\lambda_i$ of~$X =
g^{-1}\eta$~\cite{Damour}.
We could  consider a more general reference metric in $X$, or
even promote it to be dynamical in the context of bigravity.

In the simple case of a 2d space-time, we can 
solve the partial differential equations in (\ref{dini}) under the
assumption of Lorentz invariance and find that only two potentials are 
allowed~\cite{uscan}
\be
V_{\pm}= a_0 + a_1 \left( \sqrt{\l_1} \pm \sqrt{\l_2} \right) = a_0 +
a_1 \text{Tr}\left(\sqrt{X}\mid_{\pm}\right)  \,. 
\label{2dpot}
\ee
 With $\sqrt{X}\mid_{\pm}$ we denote the two possible different
 branches of the square root of $X$ in two dimensions (modulo an overall sign). The two potentials are therefore associated with the two different branches of the matrix square root\footnote{In general, an $n\times n$ matrix with $n$ distinct  non-zero eigenvalues has $2^n$ square roots, according to each of the possible choice in the sign of the square root of its eigenvalues.   Assuming that a matrix $M$ with eigenvalues $\l_i$ can be diagonalized by a matrix $U$, $M=U\,M_D\, U^{-1}$,  we have that $\sqrt{M}=U\,\sqrt{M_D}\, U^{-1}$ where $\sqrt{M_D}=\text{diag}(\pm\,\sqrt{\lambda_1}\,,\, \dots \,,\,  \pm\, \sqrt{\lambda_n})$.  We will  assume that $\sqrt{\lambda_i}$ are real and positive.}.   
 Extending this observation to  four dimensions, one can check that a large class of potentials
satisfy the conditions (\ref{dini}):
\be
V_{\text{br}}=\sum_{n=0}^3 \, \frac{a_n}{n !}\; S_n\mid_{\text{br}} \,, \label{genpot}
\ee 
where $S_n\mid_{\text{br}}$ are the symmetric polynomials of
$\sqrt{X}\mid_{\text{br}}$ once a choice of branch for the square root is made
\bea
&& S_0=1\,, \quad
S_1=\tau_1\,, \quad
S_2=\tau_1^2-\tau_2\,, \nb \\
&& S_3=\tau_1^3-3\,\tau_1\,\tau_2+2\,\tau_3\,,
\eea 
with $\tau_n=\tr\left({\sqrt{X}\mid_{\text{br}}}^n\right )$.
The subscript $\mid_{\text{br}}$ implies that different options for the square root of $X$ can be chosen, according to the $\pm$ sign in front of each of the square root of the eigenvalues.
On the other hand,  once a choice is made, the symmetric polynomials should be
constructed consistently and no mixing between the different branches
is possible without  violating  the conditions
(\ref{dini}), i.e. without reintroducing the $6^{\text{th}}$ ghost mode\footnote{Indeed one can check that,
while (\ref{dini}) are satisfied for any choice of sign among $\pm
\sqrt{\lambda_i}$, the very same equations are not satisfied when the
combination of two branches is considered. For instance, taking 
$V_{\text{br}_1}+  b\, V_{\text{br}_2} $, to have 5 dof requires $b=0$.}.

 The dRGT potential
\cite{Gabadadze:2011} corresponds only to a single 
  branch among all those in (\ref{genpot}), i.e. the branch where $\sqrt{X_D}=\text{diag}(+ \sqrt{\lambda_1}\,,\, + \sqrt{\lambda_2} \,,\, + \sqrt{\lambda_3} \,,\,  + \sqrt{\lambda_4})\equiv\sqrt{X_D}\mid_{\text{dRGT}}$.
The other choices represent \emph{different} potentials through we can
realize non-linear ghost free theories of massive gravity.

In the following we present two different approaches to study different
branches for massive gravity.  The first one is based on the construction of $\sqrt{X}$ by
using ADM variables. Alternatively, we can construct $\sqrt{X}$
perturbatively around a chosen background.


\SEC{ADM approach.} \;\;
The ADM approach joins  together the definition of $\sqrt{X}$
given in~\cite{GF} and the auxiliary variable used in~\cite{gencons} to
solve the first of (\ref{dini}).
Following~\cite{GF}, the square root can be expressed  separating the
dependence on the lapse: 
\be
\label{defX}
N\sqrt{X}=\mathbb{A} + N\,\mathbb{B} \,,
\ee
where $\mathbb{A}$ and $\mathbb{B}$ are independent from $N$. Squaring
the above equation one gets a quadratic set of equations for
$\mathbb{A}$ and $\mathbb{B}$, leading to different possible branches. 
The choice made in~\cite{GF} corresponds to the dRGT potential; on the other hand different choices are possible.
Let us show  how different branches arise and
their properties.
It is useful to introduce a new shift variable $\xi^i$ such that, in the new variables, the action becomes linear in the lapse.
The transformation is written implicitly as
\be
\label{N}
N^i = N \,\xi^i + \mathcal{Q}^i \,, 
\ee
where
$\mathcal{Q}^i=\mathcal{Q}^i(\gamma_{jk},\,\xi^j)$ will be specified
later.
Solving equation (\ref{defX}) for $\mathbb{A}$ and $\mathbb{B}$ translates in the following quadratic equations
\bea&&
\mathbb{A}^2=
\left(
\begin{array}{cc}
 1 & \mathcal{Q}^i \\
 -\mathcal{Q}^j & -\mathcal{Q}^{i}\;\mathcal{Q}^{j} \\
\end{array}
\right),\qquad 
\mathbb{B}^2= 
\left(
\begin{array}{cc}
 0 & 0 \\
 0 & \mathcal{K}_{(3\times 3)}   \\
\end{array}
\right)\,, \nb\\
&&
\mathbb{A}\;\mathbb{B}+\mathbb{B}\;\mathbb{A}= 
\left(
\begin{array}{cc}
 0 & \xi^i \\
 -\xi^j & -\mathcal{Q}^i \;\xi^j -\mathcal{Q}^j \;\xi^i   \\
\end{array}
\right) \,,
\eea
where $ \mathcal{K}\equiv \mathcal{K}^{ij}=\gamma^{ij}-\xi^i\; \xi^j$. 
We can focus for instance on two different branches for $\mathbb{A}$ parametrised as follow
\be
\mathbb{A}=\frac{\epsilon_{\mathbb{A}}}{\sqrt{1-\mathcal{Q}^k \mathcal{Q}^k}}
\left(
\begin{array}{cc}
 1 & \mathcal{Q}^i \\
 -\mathcal{Q}^j & -\mathcal{Q}^{i}\mathcal{Q}^{j} \\
\end{array}
\right),
\, \mathbb{B}=
\left(
\begin{array}{cc}
 0 & 0 \\
 0 & \sqrt{\mathcal{K}} \mid_{\text{br}}  \\
\end{array}
\right)
\ee
where $\epsilon _{\mathbb{A}}^2=1$ and
\be
\mathcal{Q}^i=\epsilon _{\mathbb A} \;
\sqrt{1-\xi^2 }   \;\;\left(\mathcal{K}^{-1/2}\mid_{\text{br}}\right)^{ij}\,\xi^j \,, \label{Q}
   \ee
with $\xi^2=\gamma_{ij}\xi^i\xi^j$.
The various branches of the square root of $X$ are controlled by
$\epsilon _{\mathbb A}$ and by the branches of $\sqrt{\mathcal{K}}$ which 
is kept formal and what follows applies to any of its
branches. Other choices of square roots are available besides this
simple one, however they typically lead to a spontaneous breaking of
rotational invariance. 

Notice that (\ref{N}) is the most general transformation
that trivializes the requirement $\text{det}(\V_{AB}) =0$
(Monge-Amp\`ere equation \cite{Fairlie:1994in}), which is the key property of potentials with 5
dof.
Such a transformation has to be invertible due to the fact
that the equations of motion already impose a constraint on the shifts
and such requirement reads
\be\label{inv}
\text{det}\left(N\;\delta_{i}^{\,j}+
\partial_{\xi^j}\mathcal{Q}^i
\right)\neq 0 \,.
\ee
Once one finds a consistent background and perturbations around it are
considered, (\ref{inv}) has to be  satisfied order by order to get a standard expansion. 

Take now a flat background for $g$.
Though both the reference metric and the background physical metric are Minkowski space, in the unitary gauge we allow the possibility that they are not aligned, namely
\be
 \bar g_{\m\n}= \text{diag}(-c^2, 1, 1, 1)  \, .
\label{minkrot}
\ee
The misalignment is measured by $c$. 
Clearly, by a coordinate transformation  one can transform
(\ref{minkrot}) in $\eta_{\mu \nu}$, however the reference metric will
not be Minkowski anymore when $c \neq 1$. Thus, even though $V$ is Lorentz
invariant, the background value of metric breaks ``spontaneously''
the Lorentz invariance $SO(3,1)$ while the rotational invariance
$SO(3)$ is preserved. 
At the background level, for the ADM variables, we have $\bar N=c$, the
spatial metric $\bar \gamma_{ij}=\delta_{ij}$ and the shifts are
zero, $\bar N^i=0$.

\vskip .3cm

\no $\bullet$ \hskip .4cm 
If  $\xi^i$ is zero at the background level $(\bar\xi^i=0)$, to preserve rotational symmetry we need
\be
 \overline{\sqrt{{\mathcal K}}}\mid_{\text{br}}\,=
 \sqrt{\boldsymbol{1}}\mid_{\text{br}}\,
= \epsilon_{\boldsymbol{1}} \, \boldsymbol{1}_3  \,,
\ee
with $\epsilon_{\boldsymbol{1}}^2=1$. Hence
\be
\overline{\sqrt{X}}=\text{diag}(\epsilon_{{\mathbb
    A}}\;c^{-1},\;\epsilon_{\boldsymbol{1}}\, 1,\epsilon_{\boldsymbol{1}}\,
1, \epsilon_{\boldsymbol{1}}\,1) \,.
\ee
Eq. (\ref{inv}) gives
\be
c+\epsilon_{\mathbb A}\, \epsilon_{\boldsymbol{1}} \neq 0 \, .
\label{invcon}
\ee
When $c=1$, Lorentz invariance is not broken and
$\epsilon_{{\mathbb A}}=\epsilon_{\boldsymbol{1}}$. Thus 
$\overline{\sqrt{X}}=\epsilon_{{\mathbb
 A}}\,\overline{\sqrt{X}}\mid_{\text{dRGT}}$ and this is just the
dRGT branch up to an irrelevant overall sign.
For $c\neq 1$, Lorentz invariance is broken,  then eq. (\ref{invcon}) is satisfied when
$\epsilon_{{\mathbb A}}=\pm\epsilon_{\boldsymbol{1}}$.  
In this case the choice of the negative sign is not the dRGT one and is non-linearly connected with the new branch:
\be
\sqrt{X_D}\mid_{\text{New}}= \epsilon_{{\mathbb A}}\, \text{diag}(-\,\sqrt{\l_1},\, \sqrt{\l_2},\, \sqrt{\l_3},\, \sqrt{\l_4})\,, \label{newsqrt}
\ee
this choice produces a genuine new potential.

\vskip .3cm

\no $\bullet$ \hskip .4cm 
If $\bar\xi^i$ is not zero, computing the inverse of ${\cal K}^{ij}$ on the background, we have
$\bar{\mathcal{Q}}^i_{\pm}=\pm\,\epsilon_{\mathbb A}\;\bar\xi^i$, thus
$\partial_{\bar\xi^j}\bar{\mathcal{Q}}^i_{\pm}=\pm\,\epsilon_{\mathbb A}\; \delta_{ij}$,
where the additional $\pm$ comes from the branches of~${\mathcal K}^{-1/2}$ in~(\ref{Q}).
From the transformation~(\ref{N}), the fact that $\bar N^i=0$ gives $(c \pm \epsilon_{\mathbb A}) =0$;
however, condition (\ref{inv}) requires that $(c \pm \epsilon_{\mathbb A}) \neq0$. 
Thus, the case with $\bar\xi^i \neq 0$ is inconsistent.
Indeed, although the transformation~(\ref{N}) is non-linearly well defined,
being perturbatively non-invertible when $\bar\xi^i \neq 0$ implies that we cannot determine $\bar\xi^i$ in terms of the background value of the old variables.
This leaves $\overline{\sqrt{X}}$ non-uniquely determined\footnote{Actually a formal, non-standard expansion for ${\sqrt{X}}$ exist, where the degeneracy of each order in perturbations is removed by the next order.
The potentials however feature the peculiar presence of square-roots of combinations of metric fluctuations in the perturbative 
expansion.
We leave the development of this method for a future work.}.

\vskip .3cm

Summarising, around a Minkowski background the only
consistent branch of the matrix $\sqrt{X}$ that preserve Lorentz 
symmetry is the usual dRGT one. On the other hand, 
 we have shown that there is also a ghost-free  branch of   massive gravity associated with 
  a flat rotationally invariant background that breaks Lorentz boosts. Following the same
  logic, other  consistent branches 
  can be constructed choosing less symmetric backgrounds.

\SEC{Perturbative expansion.} \;\;
The very same results are obtained  starting from a perturbative definition
of the square root around a background solution of (\ref{genpot}). 
To have a well defined perturbative expansion for this class of potentials,
  it is necessary that the
perturbations of the matrix square root can be expressed in terms of
the ones of the original matrix, i.e. that the 
Sylvester equation has a unique solution\footnote{From now on, we will leave understood the subscript $\mid_{\text{br}}$ for the $\sqrt{X}$ when we refer to a general branch.}
\be
\delta\,\sqrt{X} \cdot \sqrt{X} + \sqrt{X} \cdot \delta\,\sqrt{X} = \delta\, X \,. \label{sylvestereq}
\ee
It is known \cite{Sylvester}, and recently brought back to the
attention \cite{Deffayet1, Deffayet2}, that in order to 
have a unique solution for $\delta\,\sqrt{X}$ in (\ref{sylvestereq}), the spectrum of the eigenvalues of $\sqrt{X}$, $\sigma(\sqrt{X})$, and of $-\sqrt{X}$, $\sigma(-\sqrt{X})$, should not intersect, i.e. 
\be
\sigma(\sqrt{X}) \cap \sigma(-\sqrt{X}) = \varnothing \,. \label{sylvestercond}
\ee
This result  selects the  backgrounds around which a given potential can
be expanded according to the Sylvester theorem.
Notice that the case $\bar\xi^i \neq 0$ of the previous section exactly violates the condition (\ref{sylvestercond}) in the background.

For instance, if we wish to have a Minkowski
background for $g$, this implies that at the 
background level $\overline{X}$ is the identity matrix.  Therefore there
are only two possible allowed background values for
$\overline{\sqrt{X}}$ consistent with condition (\ref{sylvestercond}), namely $\overline{\sqrt{X}}\mid_{\pm}=\pm\, \boldsymbol{1}_4$.
The resulting perturbative construction reproduces to all orders the dRGT
potential\footnote{The minus case is
equivalent to the plus one, modulo a trivial redefinition of the
coefficients $a_n$.}.
Thus, dRGT
is the only potential among the  class in (\ref{genpot}) that allows a 
standard perturbative expansion around Minkowski. 
 
On the other hand, if we require for $\overline{\sqrt{X}}$ only rotational
invariance instead of the full $SO(3,1)$ invariance, new non-trivial
different branches are possibile. 
Taking as background (\ref{minkrot}), in this case $\bar X =
\text{diag}(c^{-2},1,1,1)$, hence up to an overall sign, 
we can have two different branches 
for $\overline{\sqrt{X}}$ in accordance with (\ref{sylvestercond}), namely $\overline{\sqrt{X}}\mid_{\pm}=\text{diag}(\pm\, c^{-1}, 1,1,1)$.
While the branch $\overline{\sqrt{X}}\mid_+$ is the $0$-order dRGT one, 
the branch $\overline{\sqrt{X}}\mid_-$ will lead to the new one given in (\ref{newsqrt}).

Of course, following the same lines,  we can go further and consider a
background where also the rotational invariance is broken producing new
ghost-free potentials corresponding to other branches of
(\ref{genpot}). 
Again, the idea consists of selecting  
 a background value for the metric
that removes the common eigenvalues between the spectra of $\overline{\sqrt{X}}$ and $-\overline{\sqrt{X}}$, satisfying therefore the Sylvester theorem (\ref{sylvestercond}).
We will not consider these other branches here, however they can be
potentially interesting when studying Bianchi type cosmological solutions in the context of massive gravity.

\SEC{The new potential.} \;\;
Let us now focus on the potential $V_{\text{New}}$ of (\ref{genpot}) realised through the branch (\ref{newsqrt}) (let us set $\epsilon_{{\mathbb A}}=1$).
Expanding around  (\ref{minkrot}) and setting
$g_{\mu \nu} = \bar g_{\m\n} + h_{\mu \nu}$ we get that 
all linear terms are absent (e.g. $\bar g$ is a solution of the
equations of motion) when 
\be
a_0 + 3 \, a_1 + 3\,a_2+a_3=0 \, , \quad a_1 + 2 \, a_2+a_3 =0 \, .
\ee
Notice that $c$ represents a flat direction and is not
determined by background equations. This stems from the fact that the
lapse is a Lagrange multiplier~\cite{Comelli:2011wq}. As expected, we need two tunings 
instead of the  one of  the LI case. Expanding the potential part $S_{V_{\text{New}}}$ of
the action we get
\be
S_{V_{\text{New}}}^{(2)}= \int d^4x \frac{\left(a_2 +
    a_3\right)\left(c+1\right)}{8 }  M_{pl}^2  m^2 \left[ h_{ij} h^{ij} - h_{ii}h_{jj}\right]. \label{quadact}
\ee
It should be stressed that, although from (\ref{quadact}) the limit $c \to 1$ looks regular, this limit is in fact ill defined: indeed
in the perturbative construction of $\sqrt{X}$ singularities are
encountered in this limit. This is in agreement with the result from the existence of a unique solution for the Sylvester equation and from the ADM analysis. Therefore, the limit $c \to 1$ in the $\sqrt{X}$ and in its symmetric polynomials do not commute.

Given the form of the quadratic action (\ref{quadact}), one can  
check that --  due to the lack of mass term $m_1^2$ for vectors
perturbations $h_{0i}$ --  only 2 dof are
present in the linearized theory \cite{Rubakov,dub,PRLus}. On the other hand we know that non-perturbatively 5 dof are present,
therefore 2 vector modes and a scalar one (classified according SO(3)
group) are strongly coupled around the flat background (\ref{minkrot}).
The vanishing of $m_1^2$ is rather general~\cite{PRLus} and stems from the Lorentz invariance of the potential $V$. Indeed,
following the same method used in a classic paper by Goldstone, Salam
and Weinberg \cite{GSW} (see for instance chapter 19.2 of QFT Weinberg's
book~\cite{wei}) to prove Goldstone theorem, using a multi-index
notation $g_{\mu \nu } \to g_A$, residual LI of $V$ gives
\be
\frac{\de V}{\de g_B} \, (T \cdot g)_B = 0 \,,
\ee
where $T$ is a generic generator of $SO(3,1)$. Differentiating the previous
relation with respect to $g_A$ and evaluating the result on the
solution (\ref{minkrot}) of the equations of motion, the first
derivative of $V$ vanishes and we get
\be
 \left.{\frac{\de^2 V}{\de g_A \, \de g_B}} \right|_{\bar g} \, T_{B C}  \, \bar {g}_C = 0 \,. 
\ee
Thus the mass matrix has a zero eigenvalue for each of
the generator of $SO(3,1)$ that does not annihilate $\bar{g}$. The broken
generators are precisely the 3 boost while rotations are unbroken. As
a result the $h_{0i}$ direction represents a non trivial eigenvector
with zero eigenvalue of the mass matrix and then $m_1^2=0$. 

Concerning cosmological solutions, we find that at the background
level this potential shares  the same unfortunate feature of dRGT, 
namely it only admits open FRW solutions. Also in this case however, the problem can be overcome in the context of bigravity~\cite{cosm}.
It would be interesting to check whether  the instabilities found in~\cite{defelice} and \cite{uspert}
are present when the new potential is used\footnote{Notice that in the case of the background used in \cite{defelice}, i.e. the axisymmetric Bianchi type--I metric, also other different potentials can be considered.}. Also, 
since  the vDVZ discontinuity  \cite{DIS} is absent for the background (\ref{minkrot}) that spontaneously breaks
 Lorentz boosts \cite{Rubakov},
it would be useful to check whether new exact spherically symmetric solutions exist as the ones investigated in \cite{exact}.

\SEC{Conclusions.} \;\;
Building on the general canonical analysis of massive gravity,  we have shown  
that new ghost free, Lorentz invariant massive gravity theories exist. They are associated with the different 
branches of the square root of $X= g^{-1} \eta$ that enters in the
construction of the massive gravity  potential. The dRGT potential is the unique that
produces a Lorentz invariant expansion around Minkowski
space. However, there are other branches, different
from dRGT, characterised by the fact that the expansion around flat
space breaks ``spontaneously''  the Lorentz invariance. Given the
number of parameters, a flat background exists if at least an $SO(2)$
of $SO(3,1)$ is left unbroken. 
We have discussed the explicit case where the residual group is  $SO(3)$.
For this case, symmetry arguments show that the vector
perturbations ($h_{0i}$) have vanishing mass.
It will be of great interest to study the stability of  perturbations for the cosmological solutions of this new potential. Moreover,
since this theory is expected not to exhibit the  vDVZ discontinuity, it will be 
interesting to study the features of its spherically symmetric solutions. Finally,
  it will be important  to investigate the
ghost free character of the new branches using the first order formalism.

\SEC{Acknowledgments.} \;\;
\no We thank Fawad Hassan for useful discussions on related topics.
LP and GT would like to thank
NORDITA for hospitality during the Extended Theories of Gravity 2015
workshop during which part of the work has been done. DC and LP kindly
acknowledge the Institut d'\'Etudes Scientifiques de Carg\`ese for the
hospitality, the present work was finalized during the workshop SW9
``Hot Topics in Modern Cosmology''. 
KK is supported by the UK Science and Technology Facilities Council (STFC) grants ST/K00090/1 and ST/L005573/1. GT is supported by an STFC Advanced Fellowship ST/H005498/1.

\end{document}